\documentclass[
superscriptaddress,twocolumn, preprintnumbers,amsmath,amssymb]{revtex4-1}

\usepackage{graphicx}
\usepackage{dcolumn}
\usepackage{bm}
\usepackage{soul}
\usepackage{xcolor}
\usepackage[hidelinks]{hyperref} 

\newcommand\ee{\end{equation}}
\newcommand\be{\begin{equation}}
\newcommand\eea{\end{eqnarray}}
\newcommand\bea{\begin{eqnarray}}

\newcommand{\expect}[1]{\left\langle{#1}\right\rangle}
\newcommand\di{\partial}
\newcommand\mpl{M_{\rm pl}}
\newcommand\comment[1]{}
\def\q{{\mathbf q}}
\def\x{{\mathbf x}}
\def\k{{\mathbf k}}
\def\O{{\mathcal{O}}}
\def\bq{{\bm{q}}}

\def\bx{{\bm{x}}}

\def\R{{\bm{R}}}

\def\dd{{d}}
\def\vphi{\varphi}
\usepackage[hidelinks]{hyperref} 
\comment{
\hypersetup{
colorlinks=true,
citecolor=DarkBlue,
linkcolor=DarkBlue,
urlcolor=DarkBlue,
}}

\newcommand{\ma}[1]{{\color{black} #1}}

\begin{document}

\title{A Lower Bound on Dark Matter Mass}

\author{Mustafa A. Amin}
\affiliation{Department of Physics and Astronomy, Rice University, Houston, Texas 77005, U.S.A.,}
\thanks{mustafa.a.amin@rice.edu}
\author{Mehrdad Mirbabayi}
\affiliation{International Centre for Theoretical Physics, Trieste, Italy}
\thanks{mehrdad.mirbabayi@gmail.com}

\begin{abstract}
We argue that there is a lower bound of order {$10^{-19}$} eV on dark matter mass if it is produced after inflation via a process with finite correlation length. 
We rely on non-detection of free-streaming suppression and white-noise enhancement of density perturbations as the observational inputs.
\end{abstract}
\maketitle
\noindent

\section{Introduction}
Dark matter is essential to our understanding of the cosmos -- from the astrophysical scales relevant for dwarf galaxies to the cosmological scales in the Cosmic Microwave Background (CMB) \cite{Drlica-Wagner:2022lbd}. Dark matter makes up approximately $84\%$ of the non-relativistic matter in our cosmos \cite{Planck:2018vyg}. Its detailed nature, however, is not well understood. For example, the mass or spin of dark matter particles is not known, and we have yet to confirm any non-gravitational interactions of dark matter. Furthermore, we do not have a unique formation mechanism for dark matter in the early universe.  Given the relevance of dark matter to our understanding of the cosmos, any relatively model-independent constraint on some of its microscopic properties would be valuable. In this letter, we provide such a relatively model-independent lower bound on the mass of dark matter particles.

An approximately scale invariant initial power spectrum of dark matter density fluctuations for comoving wavenumbers $k<k_{\rm obs}\sim 10\,\rm {Mpc}^{-1}$ is consistent with current observations \cite{Irsic:2017yje,Nadler:2019zrb}. We use two effects, (1) excess white noise power and (2) suppression of power due to free-steaming, to provide a relatively model-independent lower bound on the mass of the dark matter, {$m\gtrsim 10^{-19}\,\rm eV$}, assuming that the background dark matter density results from finite-momentum rather than the homogeneous oscillations of the dark matter field. The bound is independent of the nature of the field (scalar, vector, tensor etc.) and  details of the production mechanism, but assumes this field constitutes all of dark matter and interacts only gravitationally after production. With more details of the production mechanism included, the bound can be strengthened further. Our lower bound is at least 1-2 orders of magnitude stronger than that due to the finite Jeans scale in fuzzy dark matter \cite{Irsic:2017yje,Rogers:2020ltq}. It is comparable to the recent bound due to dynamical heating of stars in ultra-faint dwarf galaxies \cite{Dalal:2022rmp}. Our bound is more general, but weaker than the one of \cite{McQuinn,Feix:2020txt}, who use a model-specific version of (1) alone. Based on inferred quasar spins and hence lack of superradiance, \cite{Unal:2020jiy} also claims a stronger bound on the mass than ours. 

To demonstrate our idea, we provide a concrete example of scalar field dark matter. We set $\hbar=c=1$.

\section{White Noise}
\begin{figure}[!h]
  \includegraphics[width=2in]{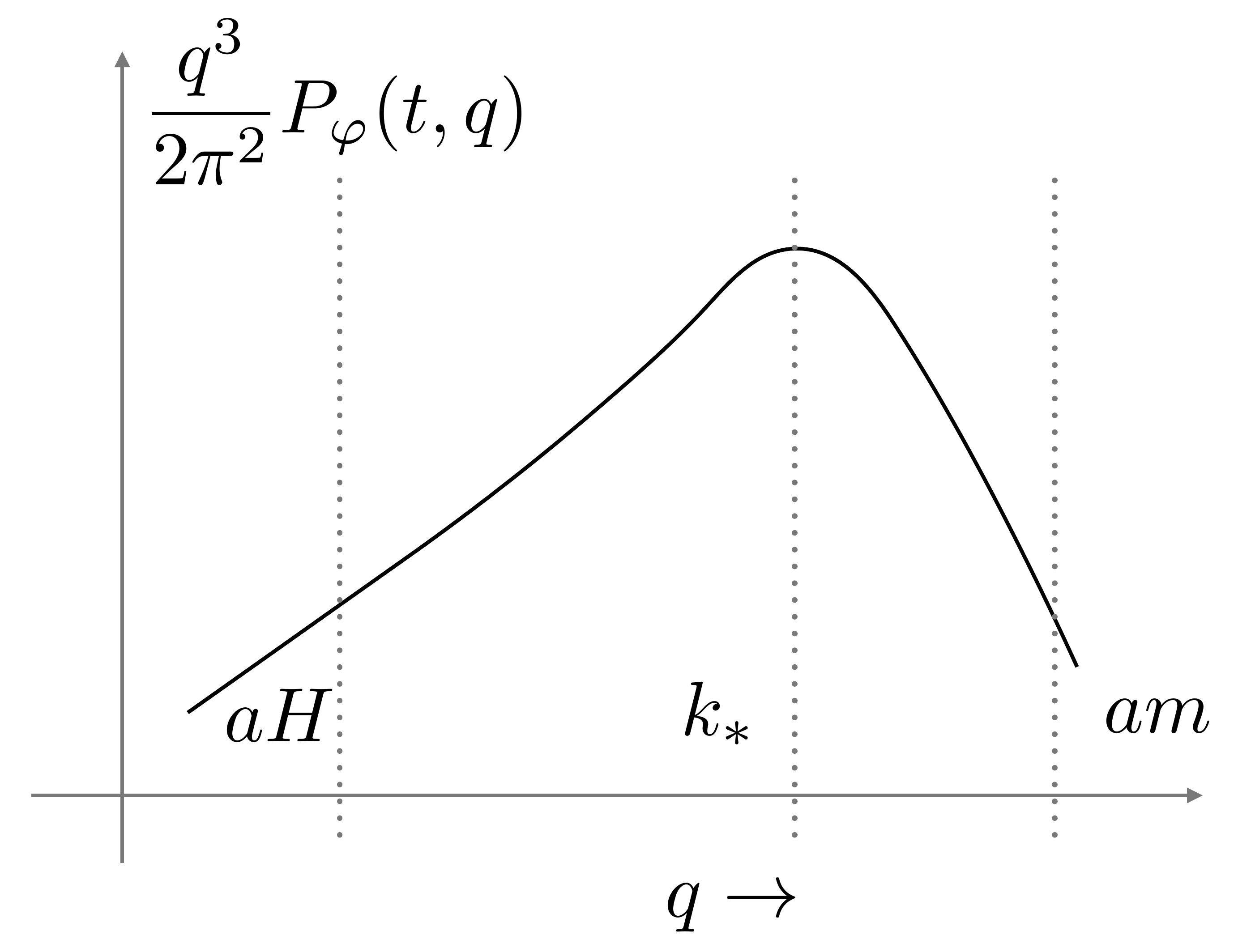}
  \caption{{Power spectrum of the field $\varphi$ at $t_i<t<t_{\rm eq}$.}}
  \label{fig:FieldPS}
\end{figure}

Consider a scalar field, $\vphi(t,\bx)$ of mass $m$, that gets excited at time $t_i$ after inflation with $H_{\rm eq}\ll m<H_i$. For now, let us neglect the inflationary adiabatic fluctuations. Then, the correlation length of the excitations is expected to be subhorizon because of causality. Near matter-radiation equality, the matter density is given by \footnote{We assume that the amplitudes of the two independent modes of the field are uncorrelated and have equal power.}
\be
\bar{\rho}(t)\approx m^2\int d\ln q\,\frac{q^3}{2\pi^2}P_\vphi(t,q)\,,
\ee
where integration over all momenta (without a UV cutoff) is a justifiable approximation because by this time the integral must be dominated by momenta much less than $m a(t)$. Meanwhile, since $H_{\rm eq}\ll H_i$, the main contribution comes from momenta much larger than $k_{\rm eq}$. For simplicity, we take it to be a single scale $k_*$ {(see Fig.~\ref{fig:FieldPS})}

\begin{figure*}
  \includegraphics[width=5.3in]{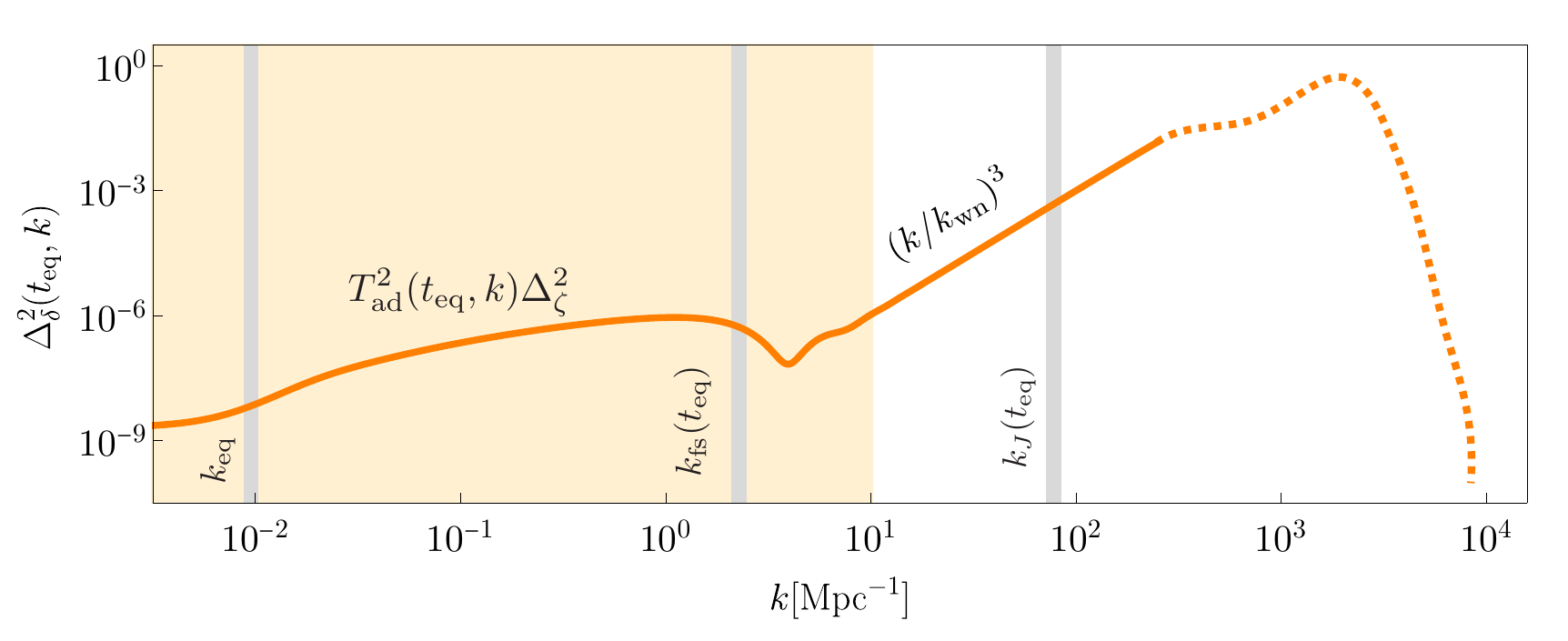}
  \caption{To illustrate our main point, we sketch the dimensionless power spectrum of dark matter density fluctuations at matter-radiation equality including the white noise excess and free-streaming cutoff. For $k_{\rm eq}<k<k_J(t)$, after equality this curve approximately shifts upwards with $a^2(t)/a_{\rm eq}^2$. Note that $k_{\rm fs}(t)\ll k_J(t)$. The orange shaded region is observationally constrained to be roughly scale-invariant. The orange curve is sketched using $m=10^{-20}\, \rm eV$ and $k_{\rm wn}\sim 10^3\,{\rm Mpc}^{-1}$-- the suppression of power due to free-streaming for $k<k_{\rm obs}\sim 10\,{\rm Mpc}^{-1}$ makes this spectrum inconsistent with observations. More generally, for the white noise contribution not to exceed the scale-invariant one at $k\lesssim k_{\rm obs}$, requires {$k_{\rm wn}\gtrsim 10^{2}\,k_{\rm obs}$}, which together with $k_{\rm fs}\gtrsim k_{\rm obs}$, leads to our lower bound: {$m\gtrsim 10^{-19}\,\rm eV$.}}
  \label{fig:PS}
\end{figure*}

 Because of the finite correlation length, at momenta $k\ll k_*$ there is a white-noise contribution to the spectrum of the fractional density perturbation $\delta$. The isocurvature transfer function is close to one and we can approximate
\begin{align}\label{wn}
P_{\delta}^{(\rm iso)}(t,k)&\approx \frac{m^4}{\bar{\rho}^2(t)}\int d\ln q\,\frac{q^3}{2\pi^2}\left[P_\vphi(t,q)\right]^2\equiv  \frac{2\pi^2}{k_{\rm wn}^3}\,.
\end{align}
$k_{\rm wn}$ is understood as being defined by the above equation. With a single scale in the problem, we expect a time-independent $k_{\rm wn} \sim k_*$.  Further details of the order unity isocurvature transfer function can be found in the supplementary material (\ref{sec:isocurvature}).

We stress that $k_{\rm wn}$ only parametrizes the slope of the white-noise part of the density power spectrum at sufficiently small $k$. It is not necessarily the location in $k$ space where the density perturbations become nonlinear. Furthermore, while not necessary for the following sections, a parameterization of $k_{\rm wn}\sim k_*$ in terms of the time and lengthscale associated with the production mechanism, and mass $m$, is provided in the supplementary material section (\ref{app:para}). 
 
 The reader who is familiar with the theory of structure formation might be skeptical about this flat spectrum. Indeed, it is well known that the stochastic contribution to the nonlinear $P_\delta(t,k)$ arising from clustering behaves as $k^4$ rather than $k^0$ at low $k$. This is a consequence of mass and momentum conservation (see \cite{Peebles}, chapter 28). A white-noise contribution $\propto k^0$, would imply that starting from the same initial matter density $\bar\rho(t_1)$, a finite-volume universe could end up with different final values of $\bar \rho(t_2)$, as a result of random clustering. Of course, this is impossible. On the other hand, it is perfectly possible that an initially radiation dominated universe ends up with different amounts of matter (i.e. different $T_{\rm eq}$) because of random fluctuations in the dark matter production scenario. For instance, there is a finite, though extremely small, probability that after Peccei-Quinn symmetry breaking, everywhere in a finite-volume universe the axion field finds itself near the bottom of the would be axion potential.

\section{Free-Streaming}
Now we include adiabatic perturbations. Initially, they modulate the energy density in $\vphi$ in the standard way, leading to the usual adiabatic contribution to the matter power spectrum at very large scales. At smaller scales, however, the subsequent evolution is non-standard due to the sizable momentum $\sim k_*$ carried by the field fluctuations. The small scale adiabatic perturbations will be washed out up to a free streaming length. During radiation epoch this length is known to grow logarithmically after $k_* <a(t) m$ \cite{Boyarsky}
\be
\label{eq:kfs1}
k^*_{\rm fs}(t)=\left[\int^t\frac{dt'}{a}\frac{(k_*/a)}{\sqrt{k_*^2/a^2+m^2}}\right]^{-1}\approx \dfrac{a^2 H m}{k_*\log\left(\frac{2am}{k_*}\right)},
\ee
where we assume that $t\lesssim t_{\rm eq}$. In the supplementary material section (\ref{sec:local}), we will see that the effect can be approximated { (at about $10\%$)} for $k\le k_{\rm fs}^*$ as a multiplicative correction to the adiabatic transfer function
\be\begin{split}\label{fs}
P_{\delta}^{(\rm ad)}(t,k)
&\approx P_{\zeta}(t_i,k) T_{\rm ad}^2(t,k)\times\\
&\left[\frac{m^2}{\bar{\rho}(t)}\int d\ln q\frac{q^3}{2\pi^2}P_\vphi(t,q)\frac{\sin[k/q_{\rm fs}(t)]}{k/q_{\rm fs}(t)}\right]^2\,
\end{split}\ee
where $q_{\rm fs}(t)\equiv \left[\int a^{-1}dt'(q/a)/\sqrt{q^2/a^2+m^2}\right]^{-1}$. 
\ma{We define a free-streaming transfer function:
\begin{align}
\label{eq:kfs2}
&T^2_{\rm fs}(t,k)\equiv\left[\frac{m^2}{\bar{\rho}(t)}\int d\ln q\frac{q^3}{2\pi^2}P_\vphi(t,q)\frac{\sin[k/q_{\rm fs}(t)]}{k/q_{\rm fs}(t)}\right]^2,\nonumber\\
&\textrm{with} \,\,T^2_{\rm fs}(t,k_{\rm hm})=1/2,
\end{align}
defining the ``half max" suppression wavenumber $k_{\rm hm}$.}

\ma{For  $q^3P_\varphi$ sharply peaked at $k_*$, we have $k_{\rm fs}^*\sim k_{\rm hm}$. As a conservative choice, we only consider free-streaming effects up to $t=t_{\rm eq}$.}\footnote{For our conservative estimates (using $k_{\rm fs}^*$) we approximate the expansion history as being $H\propto a^{-2}$ up to $a_{\rm eq}$.  Allowing for a more accurate expansion history changes the free streaming scale at $a_{\rm eq}$ by $\sim 25\%$. Furthermore, the evolution of the free streaming scale from $a_{\rm eq}$ to $a_{{\rm Ly}\alpha}\sim 0.2$ changes the free streaming scale by $\sim 15\%$.}

Famously, fuzzy dark matter has an associated Jeans scale $k_J(t)= a(t) \sqrt{m H(t)}$ above which the growth of perturbations is suppressed \cite{Hu:2000ke}. This enters $T_{\rm ad}(t,k)$, but it is not so relevant because $k_{\rm fs}(t)\ll  k_J(t)$. 
\section{Observational constraints}
We now put these two contributions together, to get a general expression for the dimensionless power spectrum 
\be\begin{split}\label{power}
\Delta_\delta^2(t,k)
&\approx T_{\rm ad}^2(t,k)\Delta^2_{\zeta}(t_i,k)\left[\frac{\sin[k/k^*_{\rm fs}(t)]}{k/k^*_{\rm fs}(t)}\right]^2\\
&+T^2_{\rm iso}(t,k)\left(\frac{k}{k_{\rm wn}}\right)^{\!3}\,.
\end{split}\ee
where $\Delta^2_f(k) \equiv k^3/(2\pi^2)P_f(k)$, and the $[\sin(k/k_{\rm fs}^*)/(k/k_{\rm fs}^*)]^2$ an approximate function that captures the behavior of $T_{\rm fs}^2$ in \eqref{eq:kfs2} for $k<k_{\rm fs}^*$. We have also included additional weak evolution of the isocurvature perturbations via the isocurvature transfer functions \footnote{Under the assumption that the two independent modes for the field have equal initial power spectra, free-streaming only effects the adiabatic part of the spectrum. Relaxing this assumption leads to an $\O(1)$ correction to the isocurvature power above the free-streaming scale.}. This model, which is valid when $k\ll k_{\rm wn}$ and  \ma{$k\lesssim k^*_{\rm fs}\sim k_{\rm hm}$}, is useful because cosmological probes are most sensitive to the onset of the new features at the smallest possible $k$. See Fig.~\ref{fig:PS} for a qualitative sketch of this power spectrum.

There are two parameters related to the microphysics of dark matter, $k_*\sim k_{\rm wn}$ and $m$, that enter this result. First, for the white noise contribution not to exceed the usual adiabatic one {($T_{\rm ad}^2(t_{\rm eq},k_{\rm obs})\Delta^2_{\zeta} \sim 10^{-6}$) when $k<k_{\rm obs}=10\,{\rm Mpc}^{-1}$, we need 
\be\label{knl0}
k_{\rm wn}\gtrsim 10^2 k_{\rm obs}\sim 10^3\,{\rm Mpc}^{-1},
\ee
a scale that re-enters the horizon at the temperature of about $0.1$ MeV.} Note the wide separation between $k_{\rm obs}$ and $k_{\rm wn}$. Second, for the free-streaming not to deplete the power spectrum significantly at $k_{\rm obs}$, we need $\ma{k_{\rm fs}^*(t_{\rm eq})}\gtrsim k_{\rm obs}$ which yields 
\be
\label{eq: bound}
m\gtrsim H_{\rm eq} \log\left(\frac{2a_{\rm eq} m}{k_{\rm wn}}\right) \frac{k_{\rm wn} k_{\rm obs}}{k_{\rm eq}^2}.
\ee
Taken together, we get {$m\gtrsim 10^{-19}$} eV. Note that we did not need to know the model dependent $k_{*}\sim k_{\rm wn}$ here, simply that it has to be larger than some value. In \cite{McQuinn}, a stronger lower bound of $3\times 10^{-17}$ eV was obtained by assuming $k_{\rm wn} \sim a(t) m$ when $H(t)= m$. Interestingly, the condition \eqref{knl0} can also be used to obtain an upper bound on dark matter mass $m < 100 M_\odot$ \cite{Murgia}.

{\em Explicit Examples.---} To further elucidate the relative model independence of our bound, we consider the following parametrized form of the field power spectra
\be
\label{eq:ParamFieldSpectra}
\frac{q^3}{2\pi^2}P_\vphi(t,q)\!=\!A(t)\!\left[\!\left(\!\frac{q}{k_*}\!\right)^{\!\!\nu}\!\!\theta(k_*-k)\!+\!\left(\!\frac{k_*}{q}\!\right)^{\!\!\alpha}\!\!\theta(k-k_*)\!\right]\!.
\ee
We take three pairs of $\{\nu,\alpha\}$ as representative examples. The $\{\nu,\alpha\}=\{3,3\}$ case is an example with a sufficiently steep ($\alpha>2$) fall-off in the field spectrum for $q>k_*$  -- typically resonant non-thermal production leads to even steeper power laws/cutoffs (c.f.~\cite{Garcia:2022vwm}). The $\{\nu,\alpha\}=\{2,1\}$ and $\{\nu,\alpha\}=\{3,1\}$ cases are motivated respectively by the inflationary production of vector dark matter scenario of \cite{Graham:2015rva},  and the axion spectrum in \cite{Vaquero} (post-inflationary Peccei-Quinn case, with strings playing an important role). These three cases yield $k_{\rm wn}\approx \{1,1.5,1.2\}k_*$ using \eqref{wn}, consistent with expectations that $k_*\sim k_{\rm wn}$. For these models, we calculate the $k_{\rm hm}$ using \eqref{eq:kfs2} with $a(t)=a_{{\rm Ly}\alpha}\approx 0.2$, and using $k_{\rm hm}\ge k_{\rm obs}$ find $m \ge \{3, 4, 4 \}\times 10^{-19}\rm eV$, consistent with our claimed bound of $m>10^{-19}\,\rm eV$.  For preliminary estimates which take advantage of existing simulations \cite{Gorghetto} specific to the axions produced by a string network, see \ref{app:QCD} in the supplementary material.\footnote{Furthermore, the well-studied warm DM scenario leads to $m\gtrsim {\rm few \times keV}$. This is because in this scenario the characteristic momentum $k_*^{\rm wdm}$ is assumed to be determined by the temperature $T\sim m$, which yields $k^{\rm wdm}_*\sim k_*\sqrt{M_{\rm pl}/m}\gg k_*$.}

\section{Summary \& Discussion}
Post-inflationary production of  light scalars \cite{McQuinn,Gorghetto_gw}, vectors \cite{Agrawal:2018vin,Co:2018lka,Dror:2018pdh,Bastero-Gil:2018uel,Long:2019lwl,Co:2021rhi}, etc. naturally leads to a combination of white-noise and a free-streaming cutoff. We have used the absence of these effects in the observational data to provide a relatively model independent and conservative lower bound on the mass of dark matter particles. Of course, observing such features in future data would be a valuable hint regarding the nature of dark matter. 

The generality of our bound arises from the following. An isocurvature white noise spectrum in the dark matter density at sufficiently small wavenumbers follows when the momentum integral that defines $P_\delta$ is dominated by field modes on subhorizon scales. This is natural for a causal production mechanism with finite correlation length. Free-streaming follows from the assumption that dark matter is non-interacting (apart from gravity). The free streaming scale depends on the mass of the particle, and is also linked to the field momentum $k_*$ that dominates the dark matter density. Our lower bound is only weakly sensitive to the details of the dark matter field power spectrum. While in some cases, the free streaming scale may also be sensitive to momenta higher than $k_*$,  that would only make our bound stronger.

Although we have assumed post-inflationary production of dark matter, our bound also applies to many scenarios where the dark matter field is produced even during inflation. More generally,  our bound is applicable when the density of dark matter is dominated by subhorizon field modes at late times, regardless of whether DM is produced during or after inflation. For example, the inflationary vector DM production scenario of \cite{Graham:2015rva} falls under the purview of our framework, so do \cite{Redi:2022llj,Harigaya:2022pjd} because they lead to subhorizon field modes dominating the energy density at late times \footnote{However, in some of these cases the abundance constraints are stronger than those from free-streaming and white noise}. Our bound is {\it not applicable} when the zero mode of the field dominates the energy density at late times --- the case for axions produced via PQ breaking {\it before} inflation.

We believe our analysis is sufficient for obtaining the quoted lower bound, however, a source of uncertainty in our bound are some of the analytic approximations we made to obtain \eqref{power} (which are quantified in the supplementary material). Obtaining more rigorous bounds for this class of dark matter models requires a new Boltzmann solver beyond, for example e.g. \cite{Blas:2011rf}, which assume a dominant homogeneous mode of the field. Due to an absence of the the zero mode of the field, the new solver will necessarily involve convolutions in Fourier space even if one only considers linearized evolution of the dark matter fields. A calculation in the nonlinear regime (including self-gravity of dark matter at sufficiently late times) will require carrying our $3+1$ numerical simulations on a lattice. These are beyond the scope of the present letter.

Finally, note that free-streaming and white noise have opposite effects on the short scale power. A more quantitative bound requires fitting the spectrum given in \eqref{power} and marginalizing over $k_*$. Note that for {$k_{\rm wn}\sim 10^2k_{\rm obs}$}, our lower bound on mass [c.f. \eqref{eq: bound}] scales quadratically with $k_{\rm obs}$. Hence, as observations push to smaller scales \cite{Valluri:2022nrh}, this bound will rapidly get stronger.
\\ \\
\ma{{\em Note added in v3}: Compared to v2, we now use $k_{\rm hm}$ defined in \eqref{eq:kfs2} for the bounds in the explicit examples section. Previously we had used a free-streaming scale defined via a Taylor expansion of \eqref{eq:kfs2} for small $k$, which leads to an overestimate of the free-streaming length when the field power spectrum has a shallow power law tail in the UV. While this did not affect the (``conservative") main bound in the paper of $10^{-19}\,\rm eV$ where we assumed that the field power spectrum has a sharp cutoff, it does affect the $\alpha=1$ cases in the explicit examples section and an appendix on axion-like particles. This has been corrected. We thank Rayne Liu, Wayne Hu and Huangyu Xiao for bringing the $\alpha=1$ case to our attention (and see their paper \cite{Liu:2024pjg}).}
\vspace{0.3cm}\\ \\
\noindent
{\em Acknowledgments.---} We thank David Dunsky, Wayne Hu, Andrew Long, Siyang Ling, Rayne Liu, Ethan Nadler, Marko Simonovi\'c, Giovanni Villadoro, Risa Wechsler, Neal Weiner, Huangyu Xiao and HongYi Zhang for helpful discussions. MA acknowledges hospitality of ICTP Trieste where part of this work was done, and is supported by a NASA grant 80NSSC20K0518.  

\bibliography{refs,reference}
\clearpage 

\section{Supplementary Material}
\subsection{Parametrization \& generalization}
\label{app:para}

It is useful to consider a fast formation process and relate $k_*$ and $k^*_{\rm fs}$ to the associated time and length scale. Suppose that the field is produced around $H_\gamma=\gamma m$, in a narrow window of wavenumbers around $k=\beta a_\gamma H_\gamma$, then $k_*\sim k_{\rm wn}\sim \beta\sqrt{\gamma}\sqrt{m/H_{\rm eq}}k_{\rm eq}$. The free-streaming suppression scale $k^*_{\rm fs}(t_{\rm eq})\sim k_*/[\beta^2\gamma\log\left(2 k_*/(\beta^2\gamma k_{\rm eq})\right)]$. The lower bound on mass, corresponds to $\beta^2\gamma\sim  10$. This is consistent with subhorizon scales dominating at production, and/or the production happening before $H=m$. Note that $\beta^2\gamma$ always appears in the same combination in $k_{\rm wn}$ and $k^*_{\rm fs}$. Also note that for $\beta\gtrsim 1$, and $\beta^2\gamma\sim 10$ , we get $a_\gamma \sim a_{\rm BBN}$, so our assumption that the universe has a standard expansion history after that point is justified. 

The results in this paper are general expectations, though exceptions can of course be found. For example, the free-streaming scale can be affected by allowing for significant non-gravitational interactions of dark matter after production or by deviating from the standard expansion history. The bound will also be relaxed if post-inflationary production accounts for part of the total dark matter. 
\subsection{Axions from a string network}\label{app:QCD}

A well motivated, but dynamically complex example of non-thermal post-inflationary production of light dark matter is that of axions produced by a string network due to Peccei-Quinn symmetry breaking after inflation \cite{Gorghetto}. In this case, there will be two widely separated momentum scales in the final field power spectrum. The larger one is the comoving scale that re-enters the horizon at PQ symmetry breaking at $f_a$. The smaller one is associated to the decay of the network, which happens when $H(t_*) \sim m_a$. It is this lower scale that dominates the number density and hence dark matter density. This would imply $k_* \sim m_a a(t_*)$. However, the numerical study of \cite{Gorghetto} suggests that for $\log(f_a/m_a)=\O(100)$ the number of strings per Hubble volume is $\xi = \O(10)$. This implies the energy density in axions at $t_*$ is much larger than the potential energy by a factor of $\xi \log(f_a/m_a)$. For a constant $m_a$, the potential starts dominating when $a(t_{\rm nr}) \sim a(t_*) (\xi \log(f_a/m_a))^{1/4}$, suggesting
\be
k_*\sim (\xi \log(f_a/m_a))^{1/4} m_a a(t_*).
\ee
We can then calculate $\ma{k_{\rm hm}(t_{{\rm Ly}\alpha})}$ \ma{from \eqref{eq:kfs2}}, using the above expression for $k_*$ (with $\xi \log(f_a/m_a)\sim 10^3$) and a model of the field spectra \eqref{eq:ParamFieldSpectra} with $\{\alpha,\nu\}=\{3,1\}$. Using $\ma{k_{\rm hm}(a_{{\rm Ly}\alpha})}\gtrsim k_{\rm obs}\sim 10\,\rm Mpc^{-1}$, then yields \ma{$m\gtrsim 2\times10^{-18}\,\rm eV$. This bound is somewhat stronger than the bound quoted in the main text for $\{\alpha,\nu\}=\{1,3\}$ ($m_a> 4\times 10^{-19}\,\rm eV$), since instead of using $k_*\gtrsim\sqrt{m_a/H_{\rm eq}}k_{\rm eq}$, we used $k_*\gtrsim 10^{3/4}\sqrt{m_a/H_{\rm eq}}k_{\rm eq}$ here.}

Note that in this concrete model, the axion mass uniquely determines $f_a$ in order to match dark matter density. Taking into account the above enhancement in $k_*$
\be
f_a \sim \mpl \left(\frac{H_{\rm eq}^2}{m_a^2 (\xi \log(f_a/m_a))^3}\right)^{1/8}.
\ee
At our lower bound of mass \ma{$m_a \sim 2\times 10^{-18}$eV, we need $f_a \sim 5\times 10^{14}$GeV}, and thus a sufficiently high scale inflation to be compatible with post-inflationary PQ symmetry breaking. Hence, there is interesting accompanying gravitational wave signatures, both primordial and secondary \cite{Gorghetto_gw}.

\subsection{Free streaming;  a simple model}
Below, we will derive formulas \eqref{wn} and \eqref{fs}. However, the main ideas can be illustrated in a much simpler example, in flat spacetime. Imagine randomly exciting a free massless scalar field at $t=0$
\be
\vphi(0,\x) = A(\x),\qquad \dot\vphi(0,\x) = \hat B(\x),
\ee
such that, in momentum space, each $A_\k$ and $B_\k \equiv \hat B_\k/k$ is an independent Gaussian variable. Their statistics are specified by
\be\label{PAB}
\expect{A_\k A_{-\k}}'= P_A(k),\qquad \expect{B_\k B_{-\k}}' = P_B(k),
\ee
where prime means a factor of $(2\pi)^3\delta^3(\k_1+\k_2)$ is dropped. In time, we have
\be
\vphi_\k(t) = A_\k \cos(kt)+ B_\k\sin(kt).
\ee
The average energy density is
\begin{align}
\expect{\rho(t,\x)}=\frac{1}{2}\expect{ \dot\vphi^2 + |\nabla\vphi|^2} 
&= \frac{1}{2}\int_\q q^2[P_A(q)+P_B(q)]\nonumber\\
&={\rm const.}
\end{align}
where $\int_\q\equiv \int d^3\q/(2\pi)^3$. Density fluctuations at momenta much smaller than those that dominate $\rho$ are given by
\begin{align}
\expect{\rho_\k(t)\rho_{-\k}(t)}'
\approx &\frac{1}{2} \int_{\q} q^4[ (P^2_A(q)+P^2_B(q)) \cos^2(\hat q\cdot \k \ t)\nonumber\\
 & +
  2P_A(q)P_B(q) \sin^2(\hat q\cdot \k \ t)],
\end{align}
where we assumed the variation of $P_A(q)$ under $q\to |\q-\k|$ is small. At short times the mixed term can be neglected and $\cos^2$ can be set to $1$. When $kt\gg 1$, we can replace $\cos^2 \approx \sin^2 \approx 1/2$. We can interpret the result as a shot-noise whose amplitude remains constant if $P_A=P_B$, while drops between $t=0$ and $t\gg 1/k$ by an $\O(1)$ factor if $P_B\neq P_A$. 

Assuming that production mechanism lasts for a time of order or longer than the peak momentum $1/k_*$, it is natural to take $P_A= P_B$. We will work under this assumption. 

To model adiabatic fluctuations, consider modulating the amplitude of the initial variations 
\be
\vphi(0,\x) =(1+\zeta(\x))^{1/2} A(\x),\quad \dot\vphi(0,\x) = (1+\zeta(\x))^{1/2}\hat B(\x),
\ee
with the same spectrum \eqref{PAB} for $A$ and $B$, and $\zeta$ is the co-moving curvature perturbation. Let's focus on a single long-wavelength adiabatic mode $\zeta(\x) = \zeta_\k e^{i\k\cdot \x}$. At time $t$, and up to $\O(\zeta)$
\be
\vphi_\q(t)\! =\! \left(A_\q + \frac{1}{2}\zeta_\k A_{\q-\k}\right)\!\cos q t +
\left(B_\q + \frac{1}{2}\zeta_\k B_{\q-\k}\right)\!\sin qt.
\ee
Up to corrections that are suppressed by $k/q$, we find
\begin{align}
\expect{\rho_\k(t)}_\zeta =&\frac{1}{2} \zeta_\k \int_\q q^2 [P_A(q)+P_B(q)]\cos(q-|\q-\k|)t)\nonumber\\
=&\frac{1}{2}  \zeta_\k \int_\q q^2 [P_A(q)+P_B(q)] \cos(\hat \q \cdot \k \ t)\nonumber\\
=& \zeta_\k \frac{\sin(kt)}{k t} \ \bar\rho.
\end{align}
This gives $\delta \rho_\k(t)/\bar \rho \approx  \zeta_\k$ when $kt\ll 1$, while it is negligible when $kt\gg 1$. This is the free-streaming cutoff.
\subsection{Cosmological perturbations}\label{sec:CosPert}
Now we analyze the evolution of perturbations during radiation dominance. Important limiting regimes are the superhorizon versus subhorizon, and relativistic versus nonrelativistic limits. At subhorizon scales, it is more convenient to work in the Newtonian gauge, where the metric perturbations remain small. In this gauge, the scalar perturbations of the metric are parametrized (at linear order) as
\be
ds^2 = - (1+ 2\Phi) dt^2 + a^2 (1+ 2\Psi) dx^2.
\ee
In the radiation era
\be
a\propto \sqrt{t}\,,
\ee
and in terms of the conformal time $d\eta = dt /a$,
\be
\Phi_\k = -\Psi_\k = -2\zeta_\k \frac{\sin(\nu\eta) - \nu \eta \cos(\nu\eta)}{\nu^3 \eta^3},
\ee
where $\nu^2=k^2/3$, and $\zeta$ is the scalar perturbation that remains conserved at superhorizon scales. One can also use a different gauge in which
\be\label{zeta}
ds^2 = -dt^2 + a^2 e^{2\zeta} dx^2.
\ee
We will use this gauge for some arguments in the upcoming section, but for most of the appendix we work in the Newtonian gauge.

In the Newtonian gauge, the equation of motion for $\vphi$, at linear order in $\Phi$, reads
\be\label{eom}\begin{split}
\frac{1}{a^3} \di_t[a^3(1-\Phi+3 \Psi) \dot\vphi]-& \frac{1}{a^2} \nabla\cdot[(1+\Phi+\Psi) \nabla \vphi]\\[10pt]
+& m^2 (1+\Phi+ 3\Psi) \vphi = 0.
\end{split}
\ee
We are treating $\vphi$ as an spectator field, namely, it feels the metric perturbations but does not source them. This is a good approximation well before matter-radiation equality.
\subsection{Adiabatic perturbations; superhorizon}
{\label{sec:AdSup}}
To set the initial condition for the subsequent evolution, we need to find the effect of superhorizon adiabatic fluctuations on fast oscillating perturbations of $\vphi$. Neglecting gradients of $\Phi$ and $\Psi$, we write the WKB solutions of eq.~\eqref{eom}:
\be\label{fplus}
f^\pm_k(t) = \frac{e^{\mp i \int^t dt' \omega(t')}}{\sqrt{2a^3(1-\Phi+3\Psi)\omega(t)}},
\ee
where
\be
\omega^2(t) \approx m^2 (1+ 2\Phi) + \frac{k^2}{a^2}(1+2 \Phi-2 \Psi).
\ee
We have ignored $\dot{H}$ and $H^2$ terms in $\omega^2(t)$. We expand
\be
\vphi(t,\x) = \int_\q [c^+_\q(\x) f_\q^+(t,\x) +c^-_\q(\x) f_\q^-(t,\x) ]e^{i\q\cdot \x},
\ee
where $c^\pm_\q(\x)$ are random variables, related by the reality of $\vphi(t,\x)$,
\be
c^+_\q(\x) = [c^-_{-\q}(\x)]^*,
\ee
and the $\x$ argument reminds us that their spectra depend on $\zeta(\x)$:
\be\label{Ppm}
\expect{c^+_\q(\x) c^+_{\q}(\x)^*}' = P(q,\zeta(\x)).
\ee
Equal and uncorrelated excitation of sine and cosine modes imply
\be\label{Ppp}
\expect{c^+_\q(\x) c^+_{-\q}(\x)}' = 0.
\ee
The energy density $u^\mu u^\nu T_{\mu \nu}$ at early times, when the field fluctuations are relativistic, is (up to linear order)
\be\label{rhor}\begin{split}
\expect{\rho_r}_\zeta &\approx \frac{1}{2}\left[(1-2\Phi) \expect{\dot\vphi^2} + \frac{(1-2\Psi)}{a^2} \expect{|\nabla\vphi|^2}\right],\\
&=\frac{1}{a^4} \int_\q q P(q,\zeta) (1- 4 \Psi(\x)).
\end{split}
\ee
In order to get the expected result for radiation $\delta \rho_r/\rho_r= 4\zeta/3$, we must have
\be\label{scale}
P(q,\zeta) = P(e^{-\zeta}q,0).
\ee
We will derive this shortly. Before doing so, we can check its consistency by evaluating $\rho$ after the fluctuations become non-relativistic:
\be
\expect{\rho_{nr}}_\zeta \approx \frac{m}{a^3} \int_\q P(q,\zeta) (1- 3 \Psi(\x))
= \bar \rho_{nr}(1+\zeta),
\ee
which is the correct relation.

Equation \eqref{scale} follows from the fact that in the presence of a superhorizon adiabatic perturbation different points on a constant $t$ slice in gauge \eqref{zeta} are at the same moment along a common history, only the $x$ coordinate has a different relation with the physical one. In particular, the energy density has to be uniform. In this gauge (and in the relativistic limit), we have
\be
\expect{\rho_{r}^{\zeta-{\rm gauge}}}_\zeta =\frac{1}{a^4} \int_\q q P(q,\zeta) (1- 4 \zeta(\x)),
\ee
where we used the appropriate WKB solution (i.e. \eqref{fplus} with $\Phi\to 0$ and $\Psi\to \zeta$). For this to be $\zeta$-independent \eqref{scale} must hold. Of course, \eqref{scale} is saying that power is the same in terms of the rescaled momentum $e^{-\zeta}k$.

Note that the WKB solutions in the $\zeta$ gauge and the Newtonian gauge are related by a time-diffeomorphism:
\be
t \to t (1- \frac{2}{3}\zeta).
\ee
Hence, the coefficients $c^\pm$ (and their spectrum) remain the same.

\subsection{Adiabatic perturbations; subhorizon}\label{sec:AdSub}
We are ultimately interested in the effect of adiabatic fluctuations with momentum $k$ on fluctuations of $\vphi$ with a much larger momentum $q\gg k$. However, we start from a global misalignment scenario, which is simpler but conceptually similar. 
\subsubsection{Inflationary/global misalignment}
Constraints on the fuzzy dark matter mass come from the lowest $k$ modes that are significantly modified compared to the CDM scenario. Hence, it is legitimate to assume $m\gg H$ when $k\sim a H$. The $\vphi$ equation can be treated with the WKB method, except the gradient and time-dependence of $\Phi$ are no longer negligible. We write
\be
\vphi(t,\x) =\vphi_0\frac{[1+ d(t,\x)]^{1/2}}{\sqrt{2a^3}} e^{-im (t - S(t,\x))} + {\rm c.c.},
\ee
where $d, \partial{S} = \O(\Phi)$. Linearizing in them, neglecting terms that are suppressed by $H^2/m^2$, and setting $\Psi =- \Phi$ gives
\be\label{delS}\begin{split}
\dot d_\k -\frac{k^2}{a^2} S_\k  - \ddot S_\k &=  4 \dot\Phi_\k, \\[10 pt]
\dot S_\k + \frac{k^2}{4 m^2 a^2} d_\k &=  -\Phi_\k.
\end{split}
\ee
When $k/a \ll H$, then $\dot\Phi \approx 0$, and we have
\be
d_\k = \zeta_\k,\qquad S_\k = \frac{2}{3} \zeta_\k t + {\rm const.}
\ee
where the integration constant was fixed by matching to the superhorizon solution. 

Away from this limit, we can eliminate $S$ to find
\be\label{eqd}
a^2\di_t \left[a^2 \di_t \left[\left(1+\frac{k^2}{4 m^2 a^2}\right) d_\k\right]\right]+ \frac{k^4}{4 m^2} d_\k
= J_\k,
\ee
where
\be\label{J}
J_\k \equiv 3a^2 \di_t( a^2\dot\Phi_\k)- a^2 k^2 \Phi_\k.
\ee
In the non-relativistic regime, the $k^2$ term in the square brackets in \eqref{eqd} can be ignored. Let us define $t_k$ as the time at which $k/a(t_k) = H(t_k)$. Explicitly, $t_k = a_{\rm eq}^2 H_{\rm eq}/(2k^2)$. In terms of the time variable
\be
\tau \equiv \frac{1}{2a_{\rm eq}^2 H_{\rm eq}} \log(t/t_k),
\ee
the non-relativistic regime corresponds to $a_{\rm eq}^2 H_{\rm eq} \tau > \log(k^2/m H_{\rm eq} a_{\rm eq}^2)$. In this regime, the solution is
\be
\label{eq:sol}
d_\k(t)\!=\! \zeta_\k \cos\!\left(\frac{k^2}{2m}\tau\!\right)
+ \int_{-\infty}^{\tau(t)} \!\!\!d\tilde\tau J_\k(\tilde\tau)
\frac{2m}{k^2} \sin\!\left[\frac{k^2}{2m}(\tau-\tilde\tau)\right].
\ee
Since the source dominates around horizon crossing, we extended the lower limit of the integral to $-\infty$. At $t\gg t_k$, the gravitational potentials becomes negligible, and we can write
\be
d_\k(t) \approx \alpha \zeta_\k
\frac{\sin\left[\beta_k\log(t/t_*)\right]}{\beta_k},\qquad 
\beta_k\equiv  \frac{k}{4m a(t_k)}
\ee
where $\alpha = 3$, and $t_* \approx 3 t_k$.  After horizon crossing, the perturbation grows logarithmically, as in the case of CDM. However, eventually the Jeans length becomes larger than the wavelength and the growth stops.

The perturbation to the phase of $\vphi(t,\x)$ follows from the first equation in \eqref{delS}. When $t\gg t_k$
\be
S_\k(t)\approx \frac{a^2}{k^2} \dot d_\k = \frac{2 \alpha \zeta_\k}{H(t_k)} \cos[\beta_k\log(t/t_*)].
\ee
Note that even for $\zeta\ll 1$, a large phase can be accumulated because $m/H(t_k)\gg 1$, though it evolves slowly when $t>t_k$. At linear order in $\zeta$, the contribution of this phase to $\rho_\k$ is of order
\be
\frac{d\rho_S}{\rho} \sim \dot S \sim \alpha\zeta \frac{H(t_k)}{m} \sin[\beta_k\log(t/t_*)],
\ee
which is negligible in our approximation. Therefore, at this order we find the expected result (with $d\rightarrow \delta$),
\be
\rho(t,\x) = \frac{m^2\vphi_0^2}{a^3} [1+d(t,\x)],
\ee
or equivalently, $\rho_\k(t)= {m^2\vphi_0^2}{a^{-3}}d_\k$ for $k\ne 0$.

\subsubsection{Post-inflation/local misalignment}\label{sec:local}
Next we consider the case where $\vphi$ fluctuations have a typical momentum $k_*$ much bigger than the momentum $k$ of the adiabatic perturbations of interest. The lowest $k$-modes that are dramatically altered in this case enter the horizon after $t_{\rm nr}$ at which $k_* = a(t_{\rm nr}) m$, i.e. $t_k\gg t_{\rm nr}$. 

We use the following WKB ansatz
\be\label{ansatz}
\vphi(t,\x) = \int_\q \vphi_\q(t,\x) e^{i\q\cdot \x},
\ee
and decompose $\vphi_\q$ in terms of positive and negative frequency modes
\begin{align}\label{vq}
  \vphi_\q(t,\x)\!\equiv\,& c^+_\q \frac{[1+ \dd^+_\q(t,\x)]^{\frac{1}{2}}}{\sqrt{2 a^3}} e^{-im \left(t + \frac{q^2}{2m^2}\int \frac{dt}{a^2}+\cdots\right)+im S^+_\q(t,\x)}\nonumber\\
  &+(+\leftrightarrow -).
  \end{align}
Note that the subscript $\q$ on the real functions $\dd^\pm_\q(t,\x)$ and $S^\pm_\q(t,\x)$ is just a label. The spatially-independent $c_\q^\pm$ satisfy the statistical properties discussed in section \ref{sec:AdSup}. We will focus on the positive frequency part and drop the superscript $+$. The negative frequency transfer function can be obtained by sending $m\to - m$ in the positive-frequency one. Neglecting terms that are suppressed by $H/m$ or $Ha/q$, we find the following system
\be\label{delS2}\begin{split}
\dot\dd_{\q,\k} + i \frac{\q\cdot\k}{m a^2}\dd_{\q,\k}
-\frac{k^2}{a^2} S_{\q,\k} - \ddot S_{\q,\k} &=  4 \dot\Phi_\k, \\[10 pt]
\dot S_{\q,\k}+i \frac{\q\cdot\k}{ma^2} S_{\q,\k}+ \frac{k^2}{4 m^2 a^2} \dd_{\q,\k}& =  -\Phi_\k.
\end{split}
\ee
In our approximation, we can replace in the first equation
\be
\ddot S_{\q,\k} \approx -\dot\Phi_\k + i \frac{\q\cdot \k}{m a^2} \Phi_\k.
\ee
Then we can eliminate $S_{\q,\k}$:
\be\label{Sq}
\!\!\!S_{\q,\k}\! =\! \frac{1}{k^2}\!\!\left[a^2 \dot\dd_{\q,\k}\!+\!i \frac{\q\cdot \k}{m} \dd_{\q,\k}\!-\!3 a^2 \dot\Phi_\k\!-\! i \frac{\q\cdot\k}{m}\Phi_\k\right]\!,
\ee
and obtain a second order equation for $\dd$, which after neglecting terms that are suppressed by $k/q$ and $q/am$, reads
\be\label{deq}
\left(a^2 \di_t + i\frac{ \q\cdot\k}{m}\right)^2 \dd_{\q,\k} + \frac{k^4}{4 m^2 }\dd_{\q,\k}
= J_{\q,\k},
\ee
where
\be\label{J2}
\!\!J_{\q,\k}\! \equiv\! 3\left(\!a^2\partial_t\!+\!i\frac{\q\cdot\k}{m}\!\right)\left(\!a^2\partial_t\!+\!i\frac{\q\cdot\k}{3m}\!\right)\Phi_\k\!- k^2a^2\Phi_\k.
\ee 

To solve \eqref{deq},  we write
\be\label{factor}
\!\!\!\!\dd_{\q,\k}(t)\! =\! \psi_{\q,\k}(t) e^{-i \k\cdot \R(\q,t)},\quad\!\! \R(\q,t)\!\equiv\! \frac{\q}{m}\int_{t_{\rm nr}}^t \!\frac{d\tilde t}{a^2(\tilde t)},\!\!
\ee
where $\psi_{\q,\k}(t)$ will now satisfy a similar equation as $\dd_\k(t)$ of the last section apart from an extra phase $e^{i\k\cdot\R(\q,t)}$ in the source. Note that the free streaming length $R(\q,t)\equiv |\R(\q,t)|$ appears naturally in the solution for density perturbations. This length grows logarithmically until $t_{\rm eq}$. Therefore, for $k\sim k_{\rm fs}$, we can neglect the extra phase in the source and the $\q\cdot\k/m$ terms in \eqref{J2} because the source contributes mainly around $t_k\ll t_{\rm eq}$, at which point $kR(\q,t_k)\ll 1$. The form of the solution is the same as \eqref{eq:sol}, and given by
\be
\psi_{\q,\k}(t)\! \approx\! \alpha_{\q,\k}\zeta_\k
\frac{\sin\left[\beta_k\log(t/t_*)\right]}{\beta_k},\quad 
\beta_k\!\equiv\!  \frac{k}{4m a(t_k)},
\ee
where $t_*\approx 3t_k$, and $\alpha_{\q,\k}= 3+ \mathcal{O}[kR(\q,t_k)]$.
 
The factorization in \eqref{factor} implies that the free-streaming suppression acts effectively as a multiplicative transfer function:
\be\label{rhopost}\begin{split}
 & \expect{\rho_\k(t)}_\zeta\\
 &\approx 3\frac{\sin\left[\beta_k\log(t/t_*)\right]}{\beta_k} \zeta_\k
  m^2\!\! \int_\q P_\vphi(t,q) \cos[\k\cdot\R(\q,t)]\\
 &=T_{\rm ad}(t,k) \zeta_\k
  m^2 \int d\log q \frac{q^3}{2\pi^2} P_\vphi(t,q) \frac{\sin[k R(\q,t)]}{k R(\q,t)}.
\end{split}
\ee
Note that the free-streaming length $R(\q,t)$ is much larger than the ``Jeans length'', because the logarithm in the argument of cosine is $\log(t/t_{\rm nr}) \gg \log(t/t_k)$. Hence, matter perturbations are still in the logarithmic growth phase when the free-streaming cutoff kicks in, and $T_{\rm ad}(t,k)$ on the second line can be replaced with the CDM transfer function. In the regime of interest, i.e. $k_*> m a(t_i)$ the above equation results in eq.~\eqref{fs} in the main text. { We expect relative corrections of order $k R(k_*,t_k)$, which for $k<k_{\rm fs}$ is $<\log(t_{k_{\rm fs}}/t_{\rm nr})/\log(t_{\rm eq}/t_{\rm nr})\sim 0.1$. }

From the first to the second line in \eqref{rhopost}, we used $\expect{\rho_\k(t)}_\zeta \approx m^2\int d^3x e^{-i\k\cdot\x}\langle\varphi^2(t,\x)\rangle$ with the form of $\varphi$ from \eqref{ansatz} and \eqref{vq}, along with our solution for $d$. To calculate the expectation value, we used the previously defined properties of $c_\q^\pm$ (see section (\ref{sec:AdSup}), but without the spatial dependence in the $c_\q^\pm$). In \eqref{rhopost}, we neglected the contribution from $S$, even though it can become large. Unlike the case of purely adiabatic perturbations, now the gradient term $a^{-2} |\nabla\vphi|^2$ can also contribute at linear order in $\zeta$. So we use \eqref{Sq} to estimate its contribution. We find
\be
\frac{\delta \rho_S}{\rho} \sim m \frac{\q \cdot \k}{a^2} S\sim \alpha_{\q,\k} \zeta \frac{H q}{m k},
\ee
which is much less than $\alpha_{\q,\k}\zeta$, when $k\gg a H$ and $q\ll a m$. \\ \\
\subsection{Isocurvature perturbations}
\label{sec:isocurvature}
Finally, suppose we neglect the small adiabatic fluctuations. Then the metric perturbations in \eqref{eom} will only be sourced by the $\vphi$ perturbations, and they are suppressed by $\rho_{\vphi}/\rho_{\rm tot}\ll 1$ deep in the radiation era. Neglecting those, $\vphi$ perturbations evolve according to \eqref{ansatz}, with $d=S = 0$. Assuming equal power in the sine and cosine modes (\eqref{Ppm} and \eqref{Ppp}), we get in the nonrelativistic limit
\be
P_\vphi(t,k) = \frac{1}{a^3} \expect{c_\k^+ c_{-\k}^-}.
\ee
Then the density power spectrum reads 
\be
\expect{\rho_\k \rho_{-\k}}' =
m^4 \int_\bq P_\vphi(t,q) P_\vphi(t,|\k-\q|).
\ee
Assuming that $P_\vphi(t,q)$ is smooth around the peak momentum $k_*\gg k$, we can expand this expression in powers of $k^2$. The zeroth order term is the leading white-noise contribution in eq.~\eqref{wn} of the main text.  Note that there is no free-streaming effect here.



\end{document}